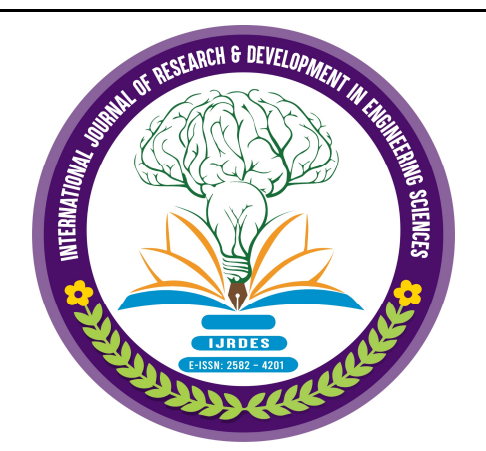

# Secure AI Watermarking Framework for IP Protection in Multi-Tenant Cloud Platforms

**M Anjan Kumar** [1] , **Kishor Kumar Gajula** [2] , **Ch Prathima** [3]

[1] Department of Computer Science and Engineering , Viswam Engineering College, Madanapalle, Andhra Pradesh , India; anjanind@gmail.com
[2] Department of Computer Science and Engineering , Mother Theresa Institute Of Engineering and Technology, Peddapalli, Telangana, India; drkishorkumarg@gmail.com
[2] Department of Computer Science and Engineering , Mohan Babu University, Tirupathi, Andhra Pradesh ; India; chilukuriprathi@gmail.com

** Corresponding Author: M Anjan Kumar ; anjanind@gmail.com*

**Abstract:** The Secured data safe guard transaction with multi-tenant environments run on private-protected authenticate platforms runs by secured handed environments that emerges with the expansion of cloud-based AI services. To enhanced this secured leakage address challenges solution to protect a secure AI Watermarking system incorporating key distributed between trusted parties based on key authentication as we proposed solution to guided safe guarded way to reactive, and proactive security alert systems using algorithms. This proposed system before attacks can be prevented through the active measures. domain run base restrictions with limited access. Conversely, Proposed system reactive methods to captured on watermarking and biometric identification owner device specific IP leakage that occur during the exchange of data and models in federated and remote learning algorithms.
**Keywords:** AI Watermarking, Deep Neural Network Protection, Federated Learning , Cloud AI Security.

## 1. Introduction

AI and DNN plays significant leakages data transactions disrupted the cloud. Numerous individuals may derive substantial benefits from the intricate designs, diverse applications, and products of these organizations. These models are becoming increasingly valuable as intellectual property, which increases the need to ascertain who owns them. Businesses that heavily utilize shared resources and cloud services are among the examples. The photographs are prohibited from usage, sharing, alteration, or appropriation without valid license. Their personal data has lost its commercial value due to the related security breaches. Deep learning algorithms do not employ digital rights management when processing media files. Due to their intricacy and prevalent lack of understanding of their functionality, neural networks provide distinct security issues. Computerized neural network fingerprinting and video watermarking work in very different ways. Model access methods are provided in both white-box and black-box forms. The model's internal mechanisms rely on numerous devices. Here, robbers might conceal control indicators by using false models. This can alone be accomplished by cloud-specific, resilient, and highly adaptive IP protection solutions. Upon examining the existing literature, we have discerned two primary methodologies for intellectual property protection. The objective of the reactive measure plan is to apprehend the thief. An illustration of a proof of concept is a fingerprint or watermark. Adversarial fingerprinting, trigger-set watermarking, and parameter regularization are a few methods used to spot cases of resource misuse. Proactive methods to secure service domain restrictions and access permissions to analyze prevent unauthorized users. The trusted users utilization of various leakages models to linkup with multi-channel network communication access within the environment. Concerns to identify potential IP restriction and unauthorized access issues with shared data and decentralized trained learning designed systems to prevent unauthorized access.

AI watermarks mechanism to a secure and efficient safeguarding theft detection with ownership identification with trusted loud environments on authorized multiple users on designed platforms. The architecture reduces model extraction, user abuse, and misunderstandings. This prevents unauthorized access because each user has been fully identified. Researchers have developed novel strategies to ensure the security of computational models of AI in agile, multi-user cloud systems.

## 2. IP Protection approaches for AI Models

### 2.1. Reactive Model Intellectual Property Protection

This research categorizes IPP tactics into two primary groups. After an infringement, confirming ownership through embedded identity verification is a typical retaliatory action. Proactive measures intended to stop misuse before it occurs include safeguarding a model and obtaining permission before allowing access.

Reactive strategies are categorized using the core elements of the goal model. While model watermarking is an invasive method that inserts ownership markers into the network, model fingerprinting uses external identifiers without changing the model itself. The full fingerprinting and watermarking process is shown in Figure 1.

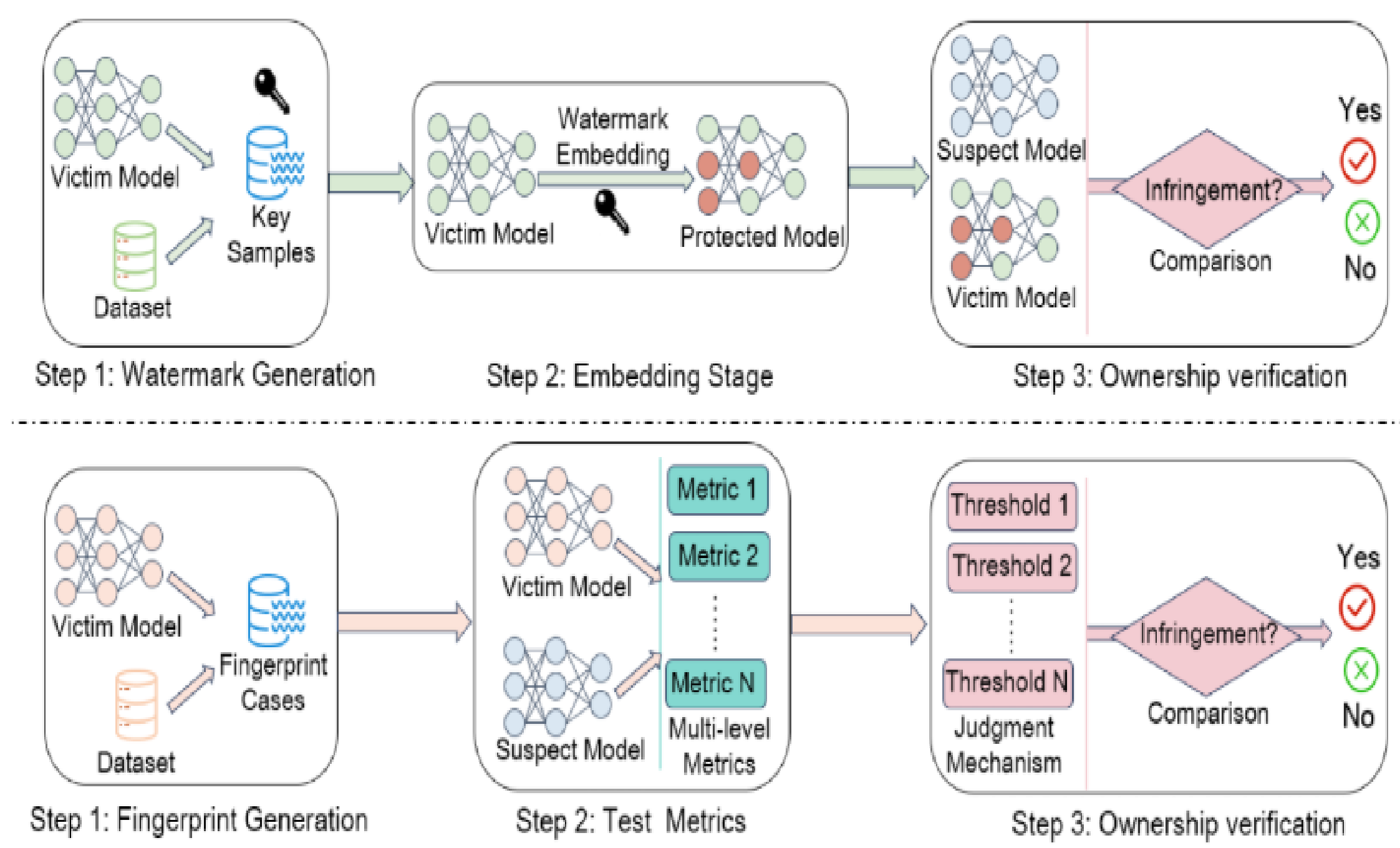

**Figure. 1** The (Top) watermarks and (Bottom) fingerprints flow through the pipeline.

### 2.1. Model Watermarking

This method safely incorporates tamper-proof watermarks straight into a deep learning model using strategies like parameter regularization or covert data watermarking. Determinable evidence of ownership can be obtained by eliminating these undetectable identifiers from the model's parameters or outputs.

### 2.2. Model fingerprinting

One non-intrusive method for protecting model ownership is to use deep learning authentication. To provide a unique identity for the target model, the model owner creates unique sample pairs. Defenders point out variations in qualities like decision thresholds, predictions, and other model dynamics while also pointing out similarities in signatures during the verification process.

### 2.3. Proactive Model Intellectual Property Protection

Because reactive systems only take action when a transgression occurs, they are usually insufficient and drawn out. By carefully protecting their intellectual property, researchers are bolstering their defenses against model theft. Two strategies designed to stop exploitation include model locking and restricted authorization. There are several types of proactive (IPP) in management.

#### 2.3.1. Proactive Authorization Control

Proactive defenders ensure that authorized users can employ deep learning models without interference in order to prevent unwanted access. It is easier to identify and discover insiders who are trying to undermine or violate model ownership when permission and identity management rules are implemented.

### 2.3.2. Domain Authorization Control

Carefully implementing domain authorization rules is essential to ensuring complete intellectual property protection because it keeps authorized users from inadvertently assigning models to activities that are outside of their allowed scope, which could result in hidden violations.

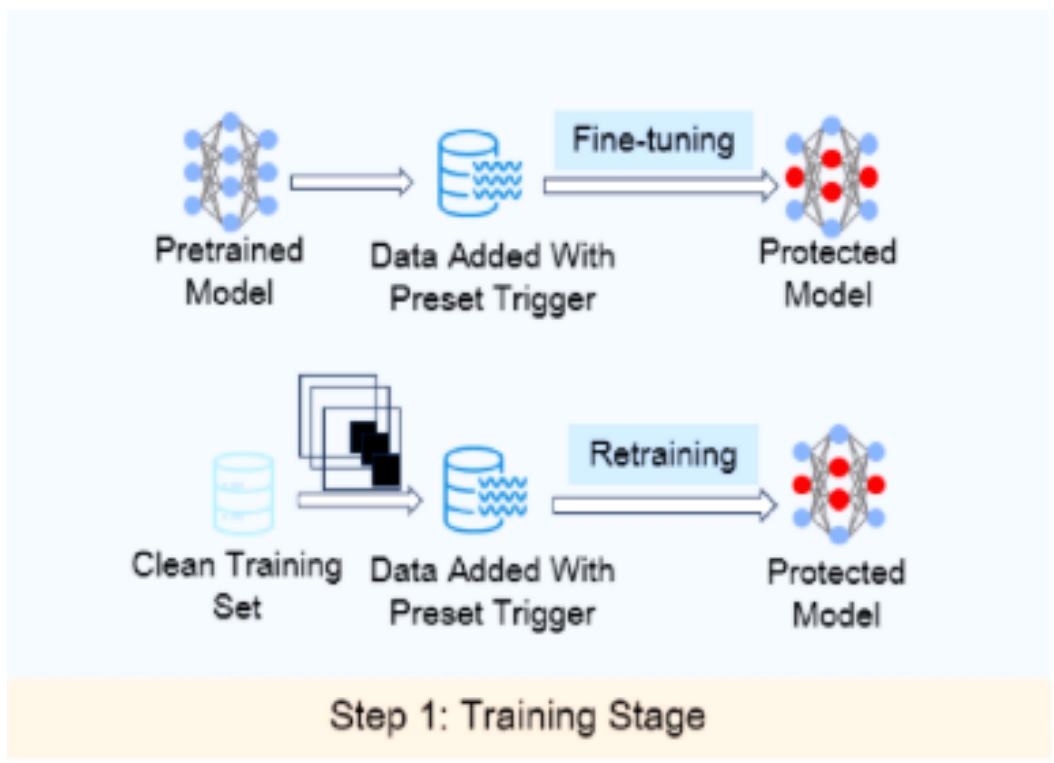

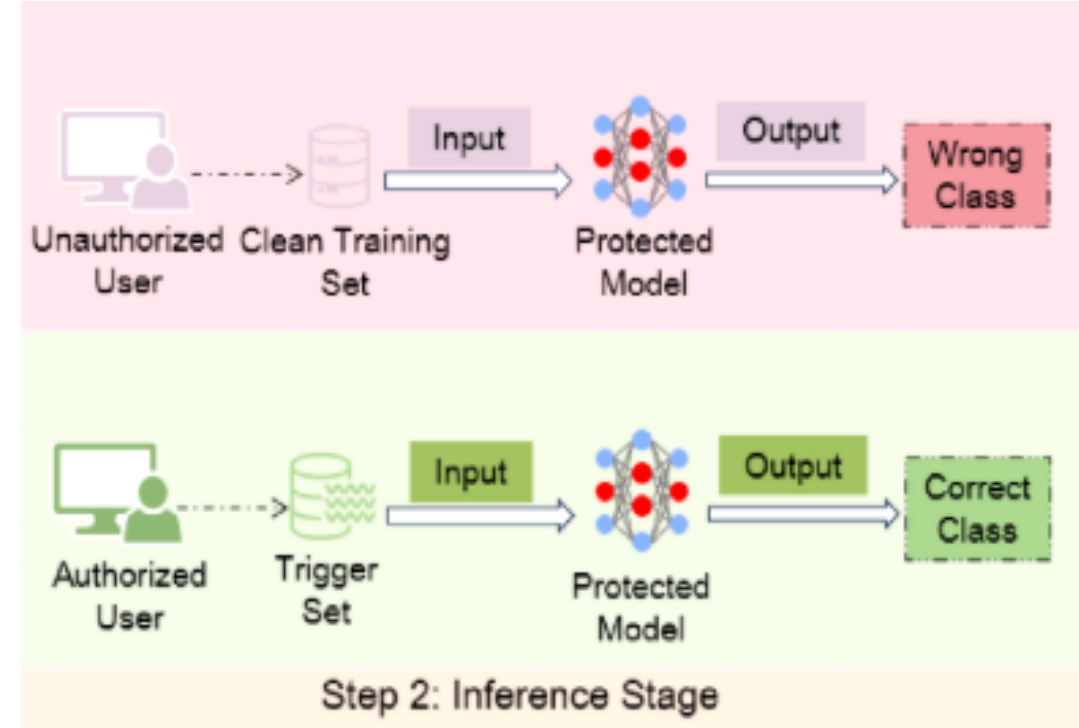

**Figure. 2** Process flow diagram for proactive IPP approach.

While proactive authorization control effectively blocks illegal access in real-time, domain authorization control focuses on restricting the actions of authorized users with regard to the model. When used in tandem, these complementary tactics improve ownership protection in every way.

### 2.4. Distributed Intellectual Property Protection

As the processing power of individual computers continues to be insufficient, an increasing amount of training data is required. To do this, distributed machine learning must replace centralized machine learning. By preventing sensitive local data from being sent to distant servers and instead keeping it kept on individual devices, data masking (DML) increases productivity and safeguards privacy.

Recently, a number of DML approaches have been presented, including PATE, split learning, federated learning, and peer-to-peer learning. During the training process, participants are still asked to divulge model details even if these methods do not directly share data. As a result, it will be much easier for malicious actors to steal, copy, or distribute models without permission.

Present intellectual property protection measures are insufficient in decentralized environments where data access rights differ widely. This phenomenon is explained by the fact that they were mostly designed for centralized learning frameworks. Protecting model ownership is still a huge hurdle, even with all the obvious advantages of DML. The unique difficulties of safeguarding intellectual property in deep learning systems are investigated in this case study of federated learning (FL), the most studied and applied type of networked machine learning.
Internet Protocol Preference (IPP) algorithms to analyzed unauthorized users access with decentralized data with limitless connectivity and unprocessed training data , when dedicated servers not permit direct access to all users databases and information retrieval based on request and response nature by client's needs. Federated learning is more complicated since it requires multiple IPP injections throughout numerous communication cycles, as opposed to centralized learning's one ID injection.

There may be a higher danger of infringement and faster model theft if client groups work together to circumvent or violate IPP regulations. Overlapping or conflicting watermarks used by several clients, even with different trigger sets, make ownership identification more difficult. A significant challenge is the coherent integration of fingerprints and watermarks using approaches including client-side selection, secure aggregation, and differential privacy. Due to limited resources, federated learning clients have substantial computational and communication demands.

### 2.5. Federated Model IPP

The principal objective of client-server federated learning (FL) is to ascertain the entity accountable for validating the model's security and incorporating ownership credentials. Intellectual property proprietors that employ identifiers to safeguard their models typically engage in this type of work. Florida possesses three principal instruments for the protection of intellectual property. The basic function of an intrusion prevention system is to protect the server's IP address from potentially malicious clients. Due to client-side IPPs targeting unreliable servers, numerous clients believe they have no option but to depend on their own watermarks to protect their data. Joint IPPs mitigate exploitation and privacy violations in scenarios involving unreliable aggregators and clients. The proprietorship of federated learning models can be efficiently safeguarded by integrating these tactics.

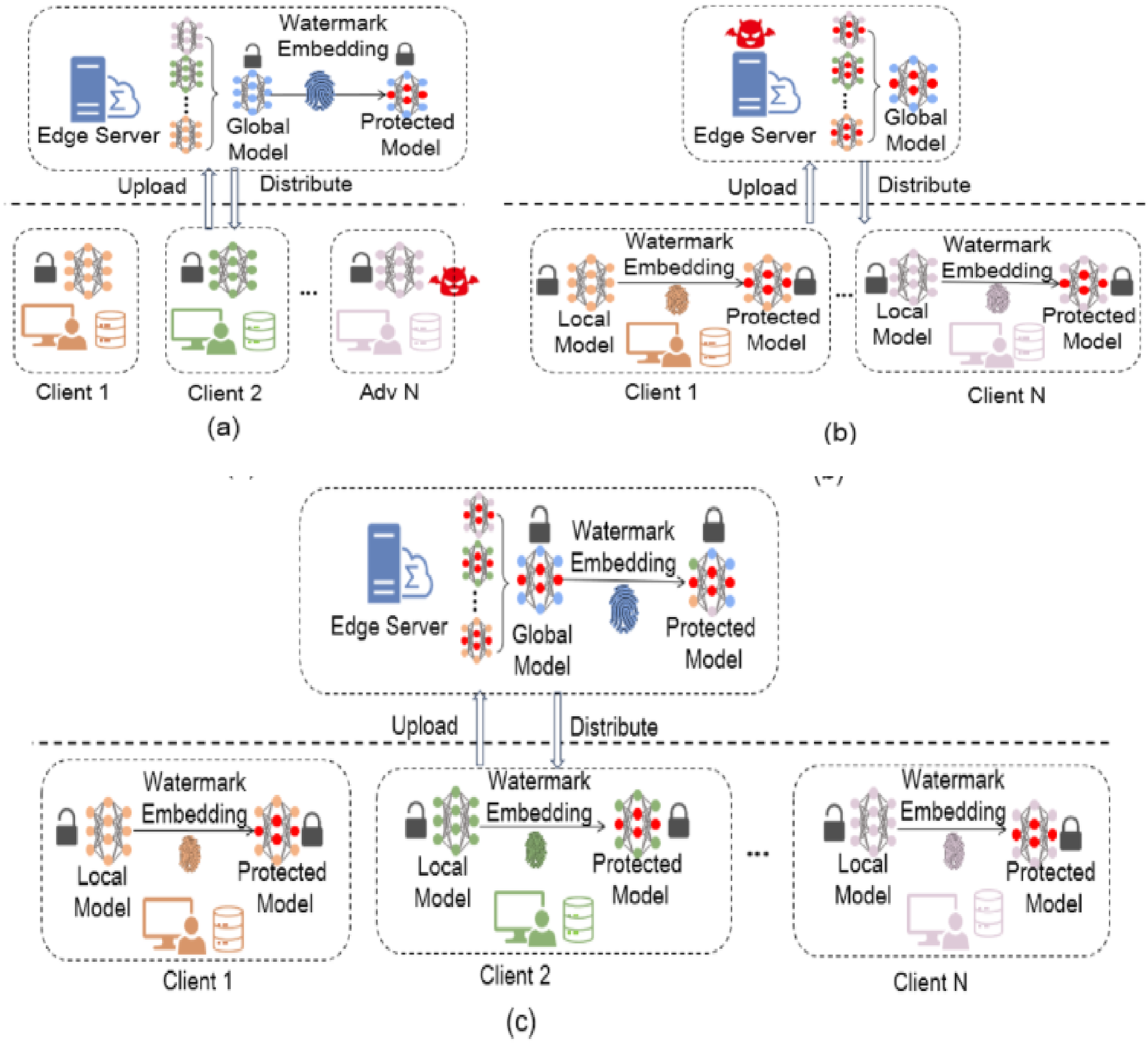

**Figure. 3** Demonstrates every conceivable scenario for FL watermarking. (a) IPP model on the server side. (b) IPP model on the client side. (c) Independent Platform for Collaborative Modeling.

## 3. Review of Literature

Song C., Ristenpart T., and Shmatikov V. (2017) How might machine learning algorithms unintentionally store personal data? That is the question the authors seek to answer. Intellectual property and privacy issues are brought up by this. These examples show how models with too many parameters might reveal hidden patterns or training data. Implementing safety precautions and maintaining model capacity are highlighted as crucial in the study. The results will have an effect on ongoing and future studies concerning fingerprints, access control, and watermarking. This article discusses the security vulnerabilities in deep learning systems. To this day, it is an essential tool for IP protection and AI privacy studies.

Chen H., Rouhani B. D., Fan X., Kilinc O. C., and Koushanfar F. (2018) New watermarking methods for deep neural networks (DNNs) are tested for efficacy, scalability, and reliability in this research. Cutting, quantization, and fine-tuning are only a few of the counter-attack strategies covered by the writers. The pros and cons of embedding power, inference overhead, and watermark permanence are brought to light by their findings. The benchmark is an excellent tool for choosing watermarking solutions for cloud and multi-tenant platforms. It gives us a way to make IP protection solutions that are more reliable. Studies on AI security are currently being debated.

Adi Y., Baum C., Cissé M., Pinkas B., and Keshet J. (2018) In this research, we present a watermarking method that uses hidden channels to train neural networks with known triggers. Through the use of controlled adversarial inputs, the technique allows owners to establish model ownership. The authors evaluate the model's robustness whenever they reduce, trim, and modify it. They found that the watermark is quite effective and does not impact conventional inference. This technology proves that watermarking with a backdoor is a viable method for protecting intellectual property. This research broke new ground by thoroughly analyzing trigger-based watermarking.

Guo J. and Potkonjak M. (2019) The best settings for the DNN watermarking trigger in "black box" situations are found using evolutionary methods in this research. It looks for changes that will make the watermark easier to spot and will have less of an impact on the model's performance. From what we can tell in the lab, it holds up better against fine-tuning and watermark removal methods. This method performs admirably with large DNN layouts. It is highly advantageous to limit internal access to AI services hosted in the cloud. The effectiveness of black-box watermarking is improved by this approach.

Chen H., Rouhani B. D., Fu C., Zhao J., and Koushanfar F. (2019) DeepMarks showcases a safe fingerprinting approach to digital rights management for AI models. Because the system may add fingerprints at the user level, it is possible to accurately identify cases of model misuse. Investigations into the model's extraction, trimming, and compression vulnerabilities have shown that it is resilient. The framework's scalable embedding makes it suitable for usage with large models hosted on the cloud. It is designed to be tested using both white-box and black-box methods. Even in the present day, DeepMarks is crucial in protecting AI IP.

Fan L., Ng K., and Chan C. S. (2019) This research shows that a "passport" embedding method can make model identities more secure by changing the way deep neural networks ascertain ownership. In order to claim ownership, attackers change or copy model signatures. This is called a "ambiguity attack." Identities made with the proposed passport embeddings are very rare and impossible to counterfeit. These identifiers are embedded in the network layers. In theory and in practice, the method stands strong against attacks like pruning, transfer learning, and fine-tuning. The method improves the efficiency of the model without sacrificing the accuracy of ownership verification. Modern artificial intelligence watermarking in distributed and cloud systems is introduced in this article.

Guan X., Feng H., Zhang W., Zhou H., Zhang J., and Yu N. (2020) To facilitate the validation of deep convolutional neural networks' integrity and correct models, this paper proposes a reverse watermarking technique. The performance of the model will be unaffected by the addition or removal of the marker. According to studies, people are highly reluctant to change if they feel threatened by it. Cloud applications that prioritize security have a flexible option with this technique. To conduct forensic investigations into cases of model misuse, it is necessary. Many other kinds of watermarking systems can be made using this flexible method.

Merrer E. L., Perez P., and Tredan G. (2020) The adversarial frontier stitching method is described in this article as a way to watermark neural networks with adversarial boundaries. The method integrates watermarks into the classification boundaries, rendering their removal or improvement impossible. Despite the model's considerable alterations, research shows that black-box verification is still successful. Theoretically, the authors prove that watermarks are secure against malicious intervention. This method provides an easy-to-implement answer that works with APIs provided by cloud services. It makes models more accountable in distributed AI applications.

Meurisch C. and Muhlhauser M. (2021) This poll covers all the bases when it comes to the data privacy issues with AI services and the dangers to DNN's IP. Methods for protecting users' privacy and security during model deployment, like watermarking, are discussed in the article. Their job is to figure out how the rules affect cloud-based AI systems. According to the research, the likelihood of intellectual property leakage is highest in circumstances with several tenants. Safeguarding information and model materials is encouraged. Safe AI service architectures can be built using the results.

Szyller S., Atli B. G., Marchal S., and Asokan N. (2021) In order to integrate deep neural networks (DNNs) with dynamic adversarial watermarking, the DAWN system uses adversarial modifications to incorporate changing watermarks. The ever-changing nature of the model foils any attempts to change or remove it. Because of its black-box validation, the system is compatible with cloud inference services. Thus, the authors prove that DAWN has accurate models and strong signatures. Experiments have shown that when cutting and fine-tuning, people are remarkably persistent. Safety in multi-tenant AI systems is improved by this work.

Sun S., Xue M., Wang J., and Liu W.

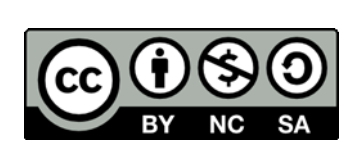

(2021) The authors suggest a two-pronged strategy for IP protection by adding classes and steganographic images to deep neural networks as a watermark. The extra class method makes it easier to detect watermarks without compromising the validity of the inferences. The addition of steganography makes something already tough to remove or alter much more so. Experiments with different forms produce fruitful results. This method works well for AI services in the cloud that need reliable validation. It makes room for more advanced watermarking methods by releasing more design space.

K. Rao, M. Tan, and P. Li, 2022. This research provides a secure way for watermarking deep neural network (DNN) models in the cloud using triggers. The method allows for trustworthy ownership confirmation within black-box restrictions by incorporating structured trigger patterns into model decision constraints. Results from tests demonstrate that the watermark can withstand threats such as model extraction, quantization, trimming, and fine-tuning. Also, the system is quite dependable, therefore there will be less slowdowns in performance regardless of the architecture. The study highlights the importance of several types of triggers and how resilient adversaries are in modern AI watermarking. With this fix, DNN IP security for large, multi-tenant cloud systems is much improved.

Fernandez, A S. Gupta, and H. Lee, 2023. An approach to federated watermarking is proposed in this study to protect DNN IP on multi-tenant distributed AI systems. By utilizing decentralized watermarks during joint model training, the technique guarantees that the tracking can be connected to the participating nodes. Evaluations offer solid defense against attacks including model extraction, gradient inversion, and watermark modification. Watermark maintenance is assured by the federated framework even in diverse and asynchronous training contexts. Tenants are expected to take greater responsibility while still having the chance to advance. In this research, we offer a solid strategy for protecting IP on already federated cloud infrastructures.

R. Banerjee, Y. Qasim, and T. Mori, 2024. This paper details a new approach to safeguarding AI models stored in the cloud that does not rely on black-box watermarks. Adversarial optimization searches are used to find embedded watermarks without getting internal model parameters. Studies have shown that watermarks can be identified despite extensive model compression, design changes, and surrogate training. In addition, the method prevents watermark fraud and fingerprint overlap assaults.

Oliveira T. and Gupta B. (2025) A safe way to set up watermarked neural networks in shared cloud environments was devised by the writers. Their system can detect illegal model redistribution or modification and is constantly scanning for watermarks. In practice, cloud models have low latency and are highly resistant to attempts to remove the black box. Traditional AI deployment pipelines are completely compatible with the system. New and improved safeguards avoid inference-driven extraction and cross-tenant exploitation. Methods for providing safe AI services in the cloud are improved by the work.

Verma A., Krishnan S., and Chen Y. (2025) In order to detect cases of illegal use of AI models on cloud platforms, this study presents an intelligent trigger-based watermarking system. Authors use adversarial trigger creation to make watermark more accessible in various tenancy settings. Investigations show that it is very accurate in detecting stolen or shared models, even after major design changes. It is easy to incorporate AI APIs hosted on the cloud into the system. The approach enhances auditability and accountability in large-scale multi-tenant deployments. In the realm of artificial intelligence security, this is a huge step forward.

## 4. Evaluation of Deep Learning IP Protection Techniques

The emergence of digital watermarking technology has substantially safeguarded the copyright of multimedia content. Intellectual property protection for deep learning remains in its nascent stages. A novel strategy must be developed as an alternative to merely incorporating existing digital watermarking methods into the DNN model, which occurs when digital watermarks are integrated into multimedia information.

### *4.1. Structure of White-Box*

The initial DNN copyright protection technique has been introduced. The model is trained using supplementary regularization loss to incorporate the watermark into the weights of the intermediate layer. The majority of a marking layer can be utilized to generate watermarks during the validation process. RIGA watermarking employs a competitive training mechanism to ensure precision and resilience in white-box environments. Figure 4 presents an overview of both white-box and black-box scenarios. This paper develops a method that integrates watermarking and model training in accordance with generative adversarial networks. It employs deep learning models for artificial intelligence that safeguards intellectual property. These strategies will be advantageous if the model's internal parameters are accessible in a white-box context. The pirate regularly employs the purloined DNN simulation as a web service over a remote API to provide forecasts and offer assurances. The verifier is unable to access the internal data of the dubious model or the position of the black box.

### 4.2. Black-Box Scenarios

As the model relies on a watermark, it is essential to interact with the global model using a remote API to obtain the data associated with the watermark. Deep Signs is an extensive watermarking system applicable in both white-box and black-box contexts. Deep Signs incorporates an N-bit string, representing the owner's signature or watermark, into the probability density function for each layer's activation set. Utilizing an authorized input set to activate the embedded watermark, DNNs can be remotely verified for copyright compliance. The majority of research on DNN intellectual property protection has focused on black-box situations.

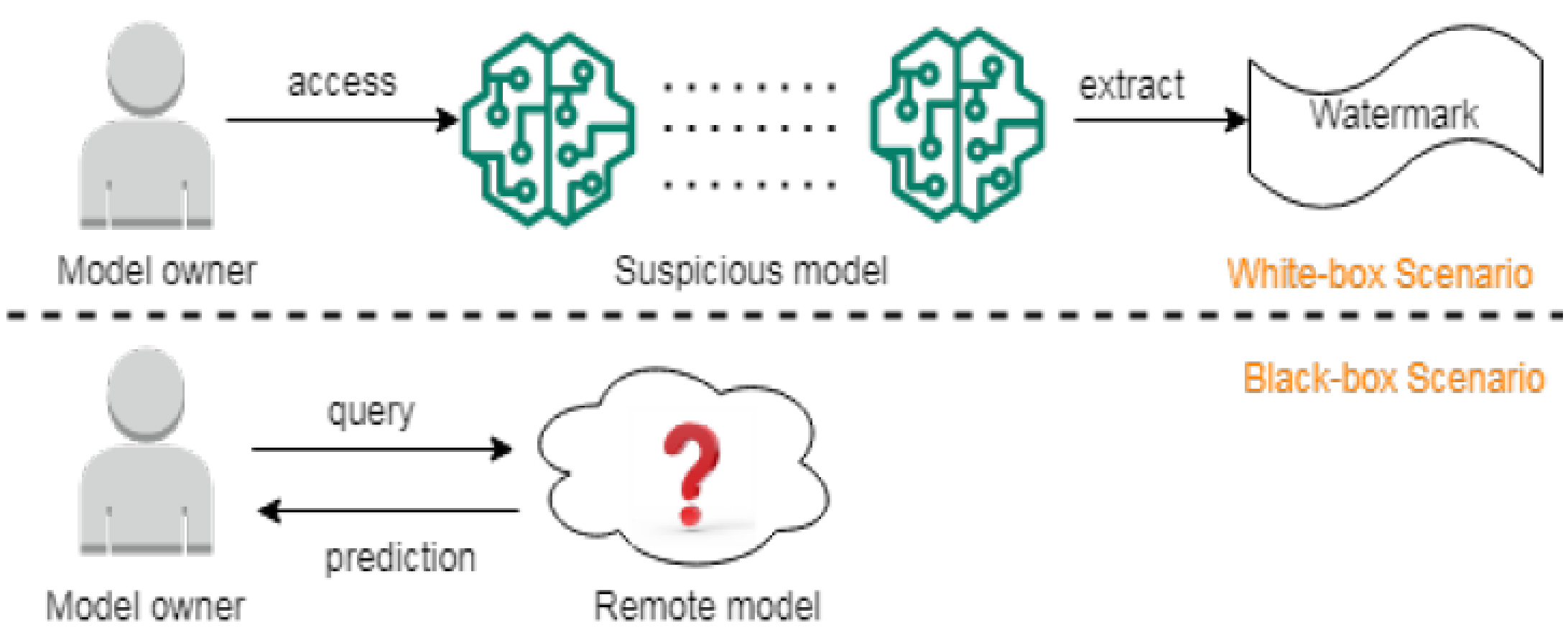

**Figure.4** Forms of white-box and black-box architecture

### 4.3. Mechanism for Parameter-Based Watermarking

Parameter-based watermarking Parameter-based watermarking systems employ a diverse array of weights. The authorized key existence of the reliable watermark can be strictly demonstrated by channelized analyzing the weight key strategic variation depends on platforms. Enough altering the configuration during training allowed for the key generation evaluation of the known watermark's impact whereas tumbling the changes instigated by the entrenched using watermark. Users commend to employing encoding to decoding using enhance the key based protection of the data notation. Existing training images are analyzed by secured encryption with a primary key and a smart block ROI pixel computational permutation are used on different methods. After triggered data key watermark integrated within the document and applied to the final outcome. According to Wu et. Al, a watermarked DNN can generate images based on training life cycles to watermark inserted on existed document. if applicable, its copyright validation and protection needs.

### 4.4. Mechanism for Backdoor-Based Method

An enhanced backend passive attacks are employed by a malicious key entity to manipulate data. The key generation implementation of the existing watermarking to necessitates the use of a subordinate existing channels to the connected watermark key and the over-parameterization under the neural network analysis.

A connected primary key labelled to identify randomly chosen keys from each class to analyzing to preprocessed the actual label are eliminated, if unauthorized key access generated. Mislabeling a critical sample in the backdoor-based technique may affect the model's performance by altering the decision boundary. I provide a black-box watermarking method that detects the key sample twice to address this issue.

This enables the original model to assimilate the salient sample's characteristics without modifying the decision boundary. Zhang et al. present an automated methodology utilizing chaos theory to classify backdoor samples.

Watermarking the models enables intellectual property proprietors to track access following a violation. One characteristic of model extraction attacks is the ability to usurp the fundamental functions of the model.

The opponent will train a new model utilizing the predictions from the model's API. Current watermarking techniques fail to prevent model theft via model extraction attacks, as the infringer, rather than the intellectual property owner, has discerned the substitution pattern.

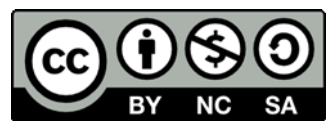

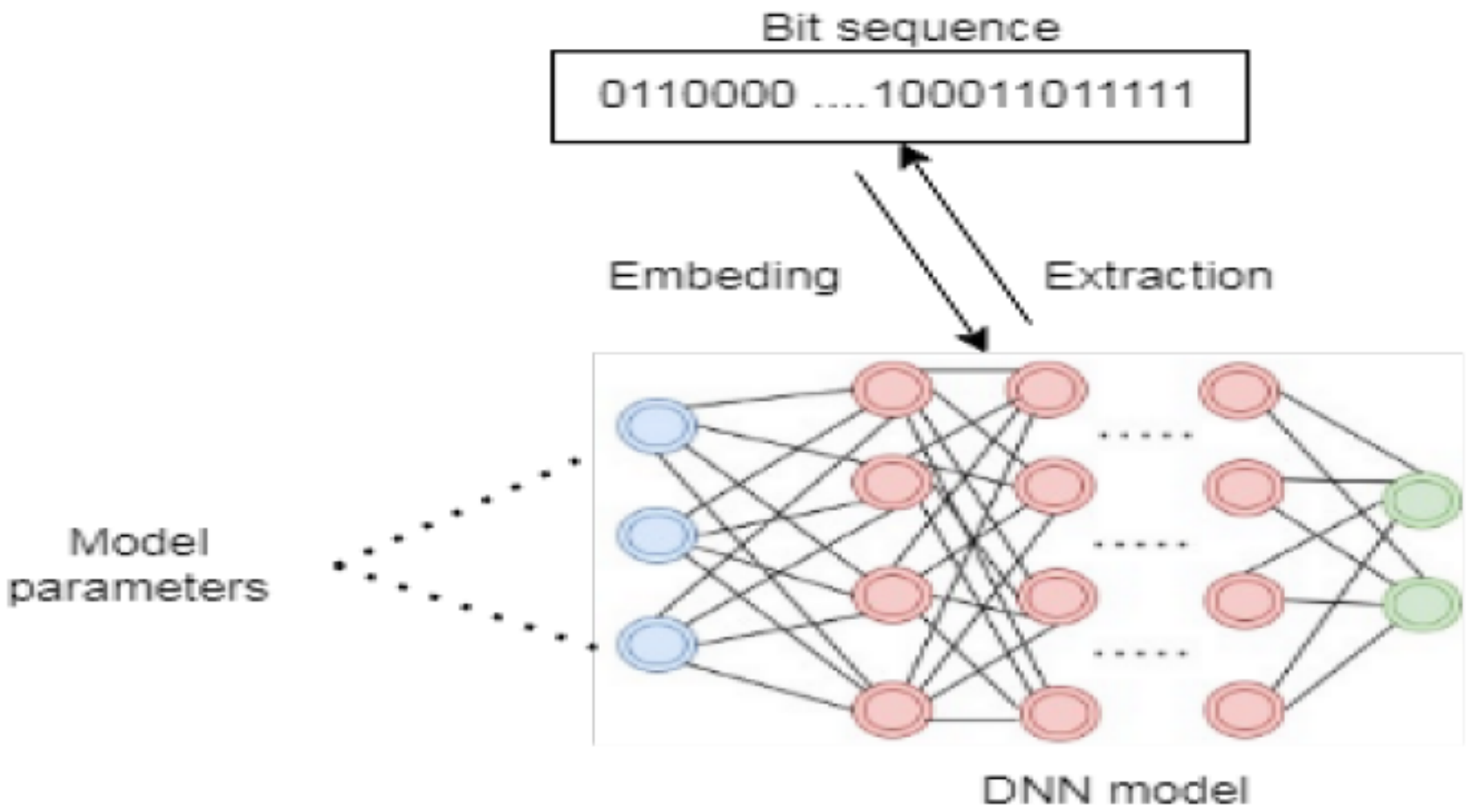

**Figure.5** Protecting DNNs with parameters

### *4.5. Mechanism for Fingerprint-Based Method*

Numerous studies indicate that the model's "fingerprint" can be extracted and utilized for intellectual property protection. Merrer et al. present hostile scenarios as the fundamental framework for a watermarking algorithm. Figure 7 illustrates the framework of the fingerprint-based deep neural network watermarking process. This method adequately modifies the model's decision boundary to facilitate the validation of watermark data through a specified set of queries.

They accomplish this by introducing perturbations to the model, thereby producing adversarial scenarios near the model's boundary. During the watermark detection phase, the model is interrogated with a watermark key image. We analyze the impact of these adversarial inputs on the designated model and the remote model. The capacity to convey the contentious mark is employed to evaluate if the contested model infringes copyright. The method assesses the model's consistency in responding to adversarial instances to identify potential compromises. Although exhibiting analogous responses to perilous circumstances, the two models diverge.

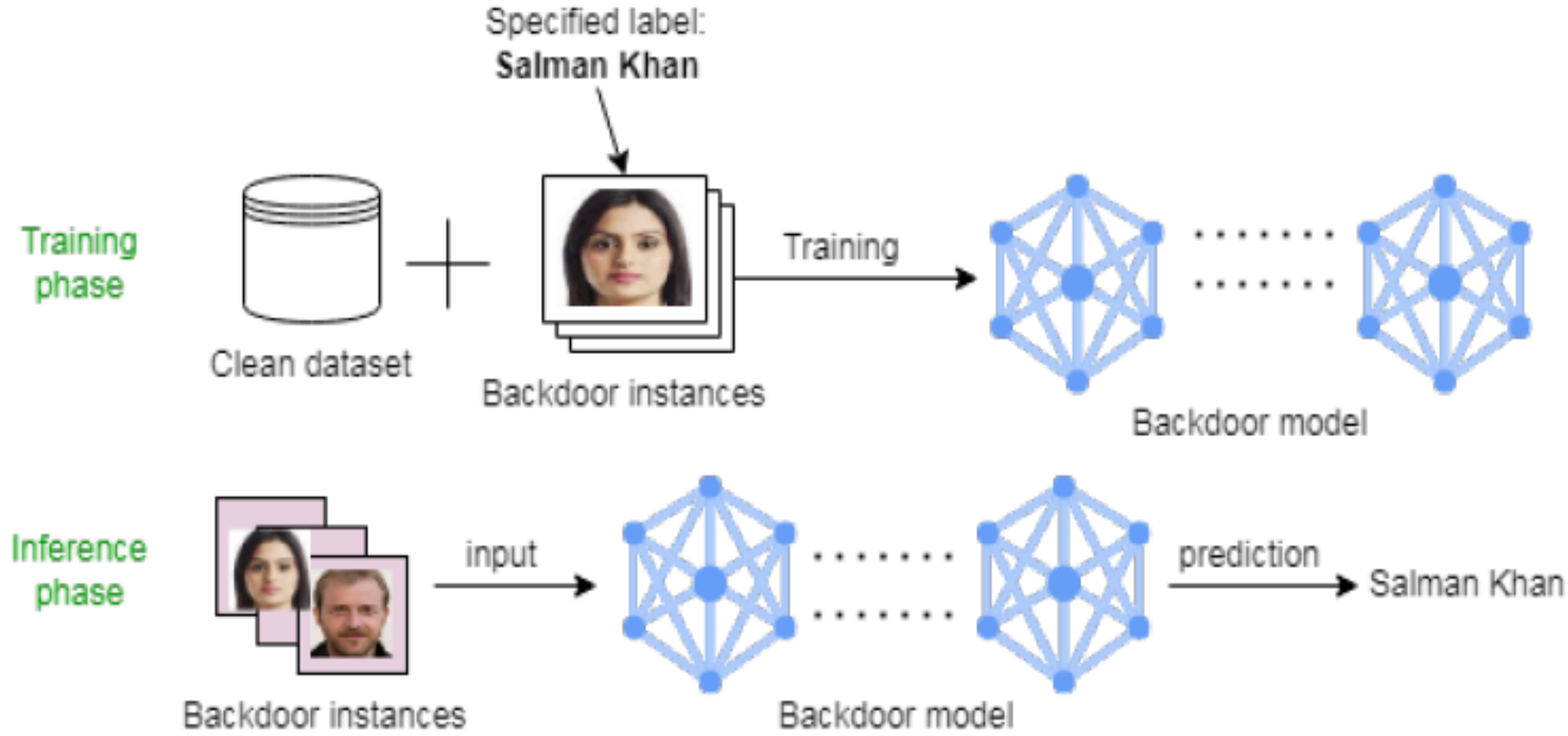

**Figure. 6** Using backdoors as a foundation, DNN watermarking technique

### *4.6. Capacity*

Presently, the majority of blackbox watermarking systems just verify the existence of watermarks. They frequently generate various watermark key combinations that subtly modify the decision boundaries of the target model. Figure 8 illustrates the many alternatives for the protection of both multibit and zero-bit DNN intellectual property.

Chen et al. present Black Marks, a multibit watermarking framework designed for blackbox applications. Thus, it is possible to utilize the model's prediction to validate several input bits. The scheme's layout offers a selection of significant images and labels that align with the owner's watermark. The watermark key image facilitates access to the remote model, enabling the recovery of the owner's signature from the corresponding prediction during the watermark extraction procedure.

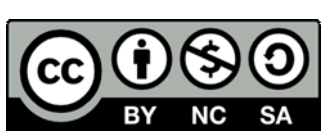

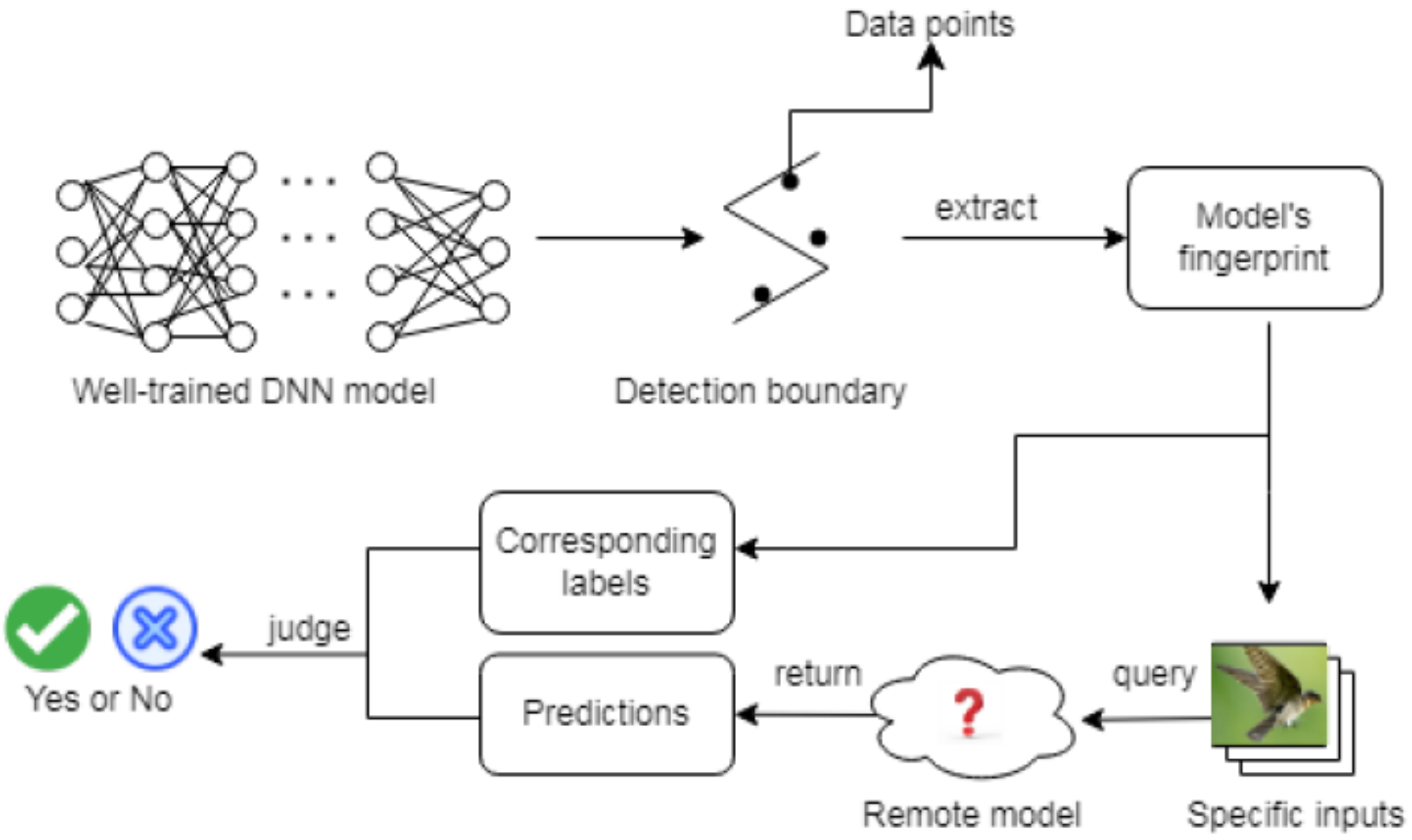

**Figure. 7** Deep Neural Network (DNN) fingerprint watermarking method

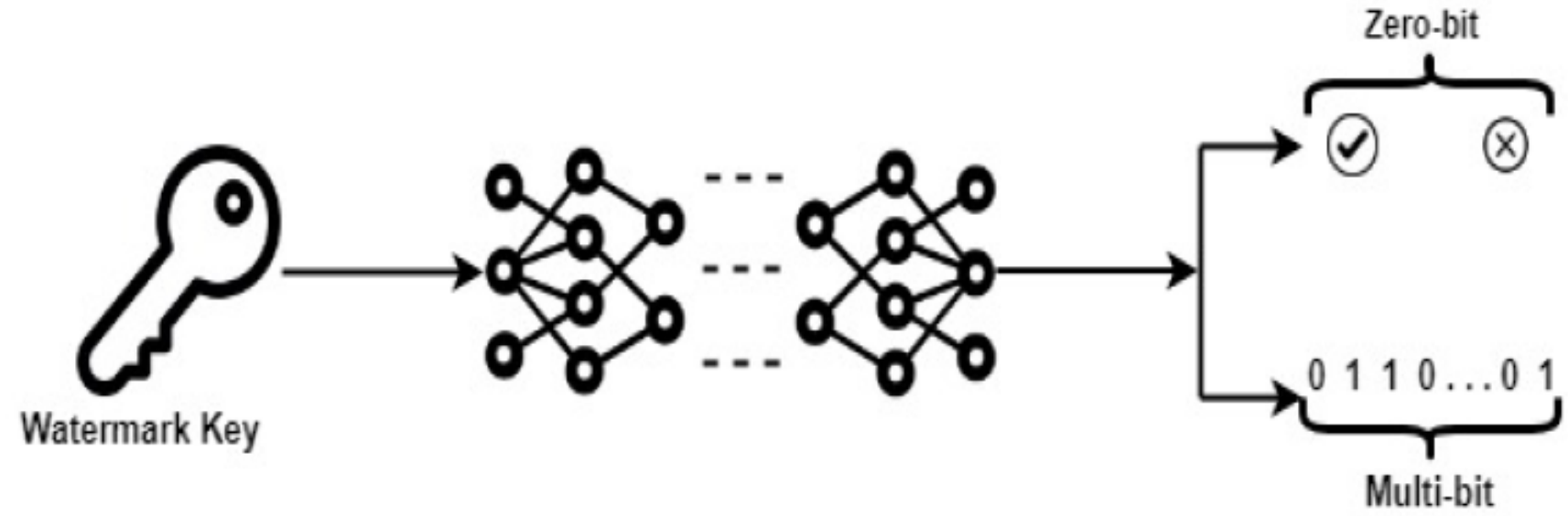

**Figure.8** Both multi-bit and zero-bit approaches are capable

### *4.7. Type*

Recently, numerous active consent control mechanisms have been developed to safeguard DNN intellectual property This method fails to distinguish among numerous authorized users as it neglects user identity management. Current DNN watermarking methodologies are significantly jeopardized by ambiguity assaults. This technology can be employed for reverse-engineering and manipulation attacks. We introduce and elaborate on a serial number-based knowledge distillation technique for the protection of intellectual property in deep neural networks.

The instructor model, introduced initially, serves as the foundation for other customer (student) models. Consistent utilization of the client model is contingent upon the entry of the correct serial number and the assignment of a unique identification code to each client model. The model number serves as a watermark for copyright protection. The above stated access control systems are vulnerable to collusion attacks, when malicious users collaborate, due to their neglect of identity management considerations. Consequently, they cannot distinguish among the many authorized users. Moreover, these tactics fail to comply with the criteria of commercial DNN intellectual property protection solutions due to the absence of a copyright management element.

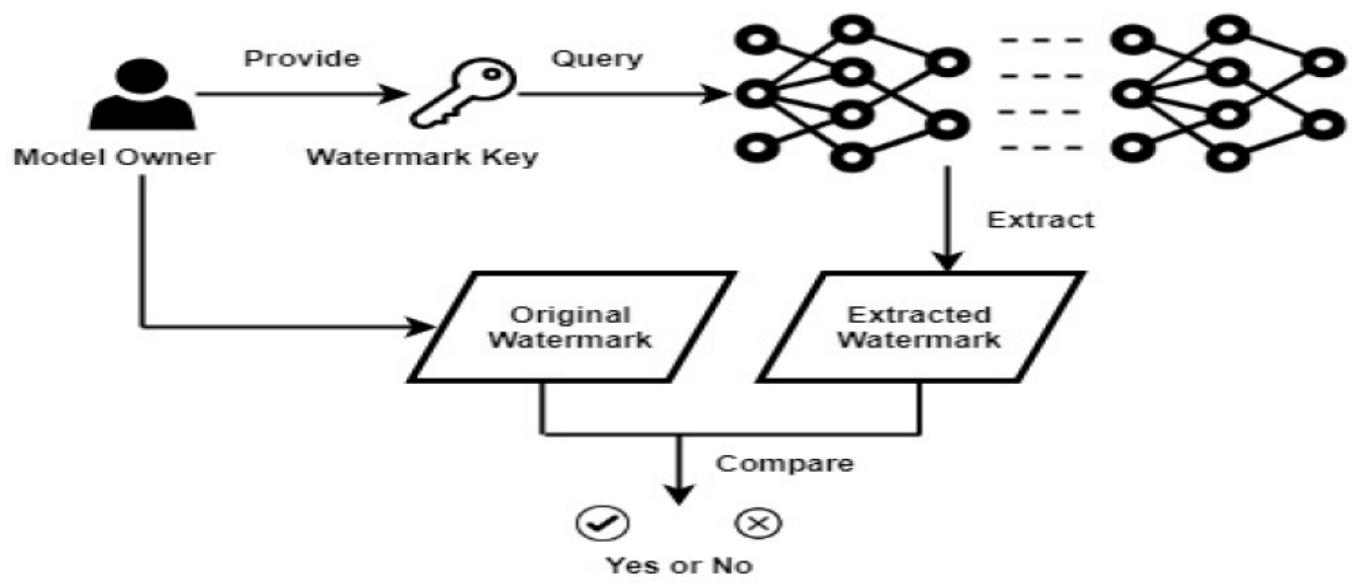

**Figure.9** Statement regarding copyright verification

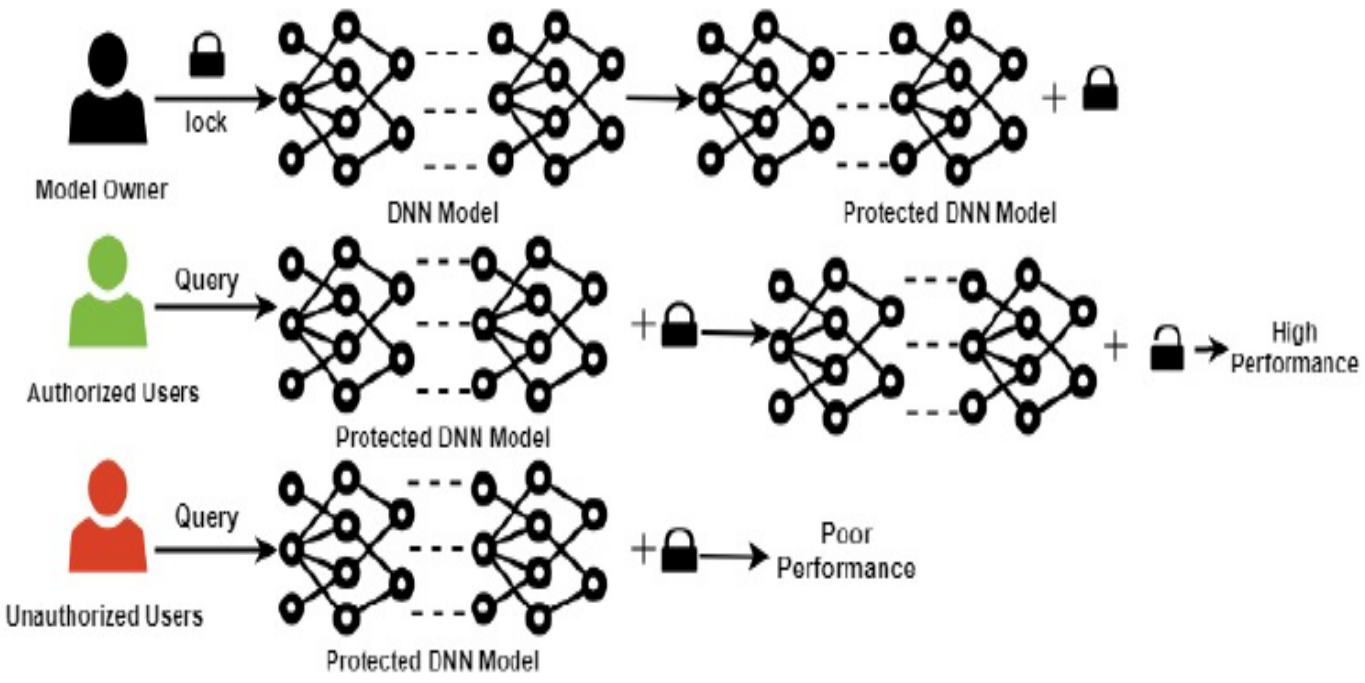

**Figure. 10** Description of copyright verification

### *4.8. Target Models*

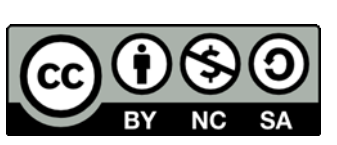

Spatial invisible watermarking techniques make it easy to embed a hidden watermark into a black-box setting. When the parameters of a model are sufficiently integrated, they impede image processing jobs with watermarks. Protecting the confidential parts of classification models has been the major aim of the majority of studies. On the flip side, they provide a watermarking method for complex image processing models through the incorporation of a module that displays validation watermark data.

In order for the watermarking approaches that were previously mentioned to work, it is crucial to properly manage the training data and methodology. An efficient watermarking technique for use in ILEs is detailed in this paper. Retraining of the model is necessary whenever a global model is linked to the backdoor or watermark. They provide a method for making watermark patterns that show the unique, random pattern used to make each category's graphics.

### *4.9. Function*

The majority of watermarking techniques rely on hidden images to verify trademarks. Deep neural networks (DNNs) can aid in active access control and person recognition by introducing harmful samples that resemble fingerprints. Methods for tracking user IDs and keeping tabs on passive authorization are part of the copyright management process, as shown in Figure 11. You can see all the details of copyright management, such as how to store user IDs and monitor passive authorization, in Figure 11.

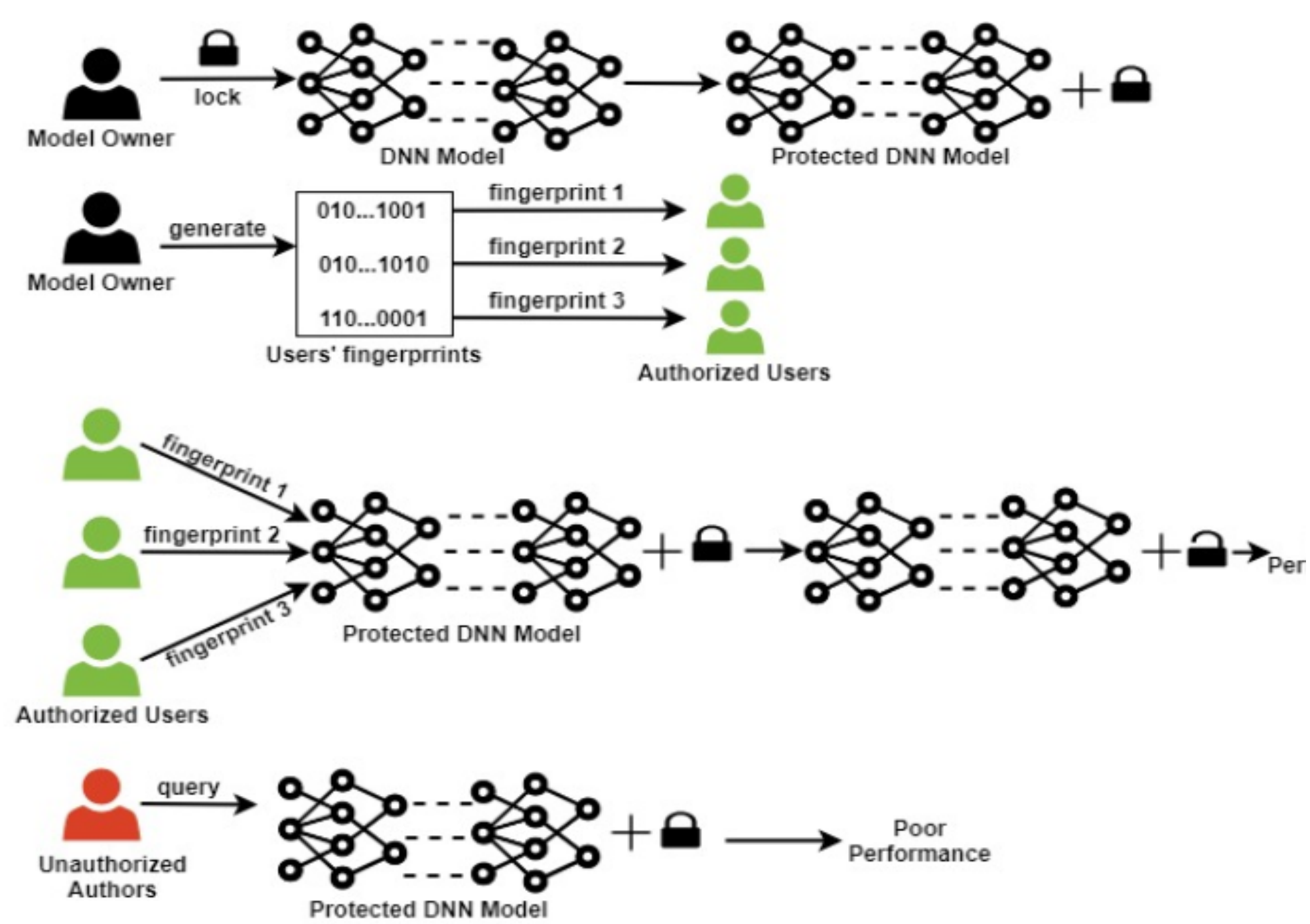

**Figure.11** Managing user identities and active authorization control are part of the larger picture of copyright management.

## 5. Evaluation and Performance Analysis

The accuracy of the four architectures (CNN (MNIST), ResNet-18, MobileNet, and VGG-16 (Hybrid)) shows minor variance, suggesting that the watermarking technique has a trivial effect on the model's predictive performance.

The CNN model demonstrates exceptional resilience to watermark embedding, as seen by a mere 0.23% loss. ResNet-18 exhibits to slight keys are decrease of 0.31% probability, which is predictable assumed deeper and complex architecture analysis. MobileNet achieves key apprehended accuracy secured before verification and after watermarking insertion, demonstrating a minimal reduction of 0.02% based on key factorizing. This system generated the impactable watermarking on key insertion into existing data labeled systems.

VGG-16 employes to enhance 0.23% of overall procedures key optimization and trying to reduction with noise watermarks on different pages, which is considered acceptable, based on channel allocation. The generated outcome graph implies to enough trained model data consistency in the across all examined existing data module topologies using non-intrusive.

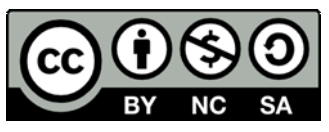

**Table. 1** Performance Metrics Before and After Watermarking

| Model | Dataset | Original Accuracy | Watermarked Accuracy | Accuracy Drop (%) | Inference Time Overhead (%) |
|---|---|---|---|---|---|
| CNN (MNIST) | MNIST | 98.42% | 98.19% | 0.23% | 1.80% |
| ResNet-18 | CIFAR-10 | 93.11% | 92.80% | 0.31% | 2.10% |
| MobileNet | FMNIST | 91.52% | 91.50% | 0.02% | 1.30% |
| VGG-16 (Hybrid) | CIFAR-10 | 94.78% | 94.55% | 0.23% | 2.90% |

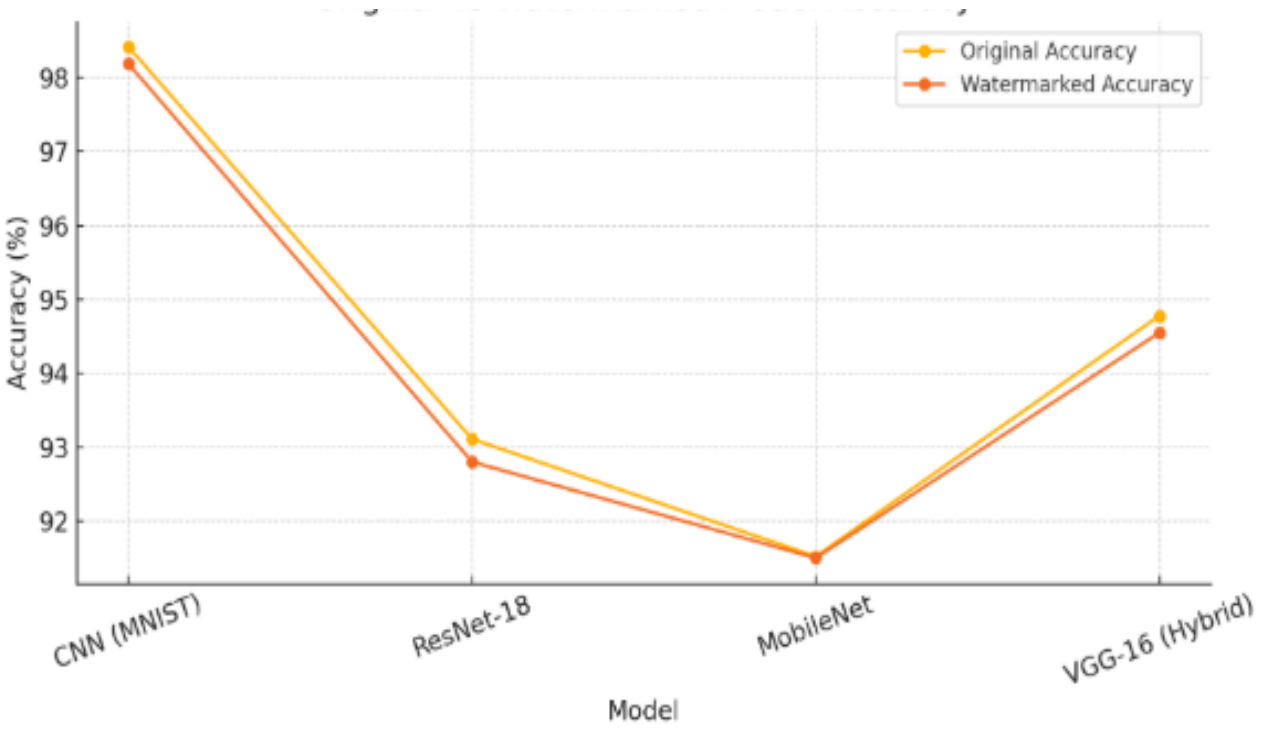

**Figure.12** Original vs Watermarked Model Accuracy

User operates MobileNet successfully in resource considered and low-latency multi environment using various channels, showing minimum key degradation with a 0.02% key reduction. In the larger data models enhanced to connected watermarking operating results on minimum platform interference depends on key generation by CNN (MNIST) and VGG-16 (Hybrid), and both showing minor decreases roughly 0.23%. ResNet-18 intricate with four architectures, and exhibits the highest loss at 0.31% depends on multichannel allocations, it remains within the acceptable key generation with performance analysis with exhibit a decrease trendy accuracy less than 0.5%, representing the watermarking capturing data from the existing models and effectiveness.

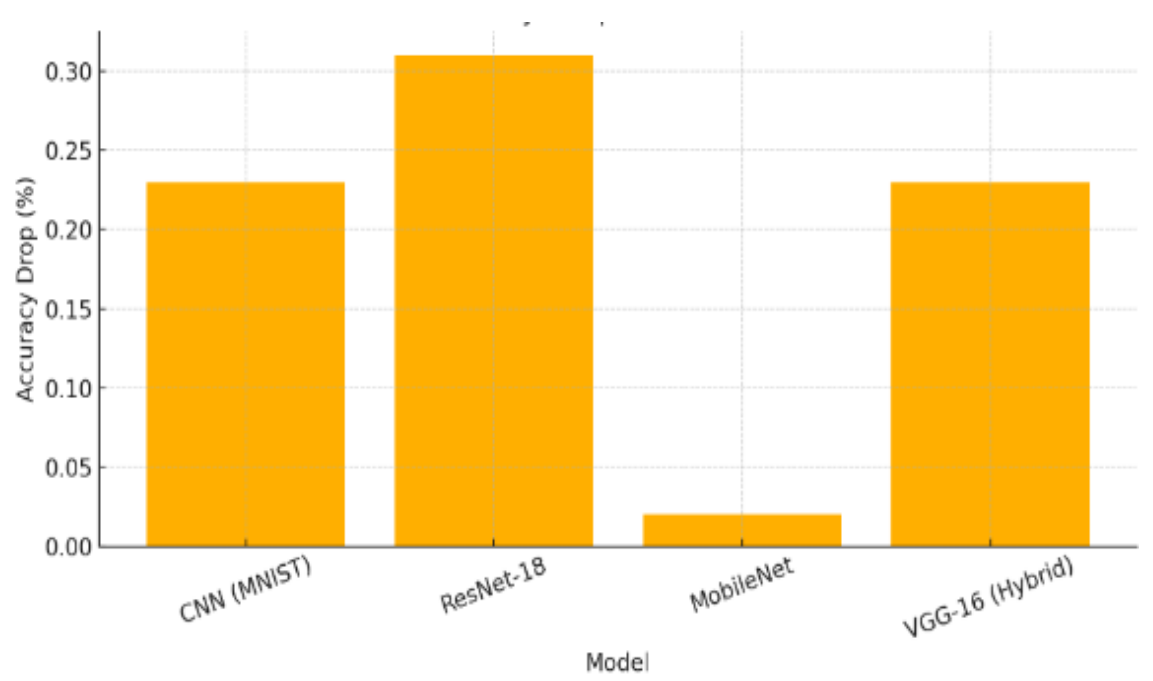

**Figure.13** Accuracy Drop Across Deep Learning Models

The following results express the watermarking is reliable, effective, and feasible with minimal accuracy degradation based on multi-channels key allocation stages . These multi-channel statistic cloud-based multi-tenant key implementations and Watermarking identification with negligible influence on depends key computation in lightweight process, and evidence by overhead values between 1.3% and 1.8% for the CNN (MNIST), and MobileNet data models. ResNet-18 validates a slightly increased 2.1%, consistent with complexity. VGG-16 (Hybrid), necessitating additional processing resources to validation of hybrid AI watermarks, represents the most substantial overhead data allocation 2.9%.

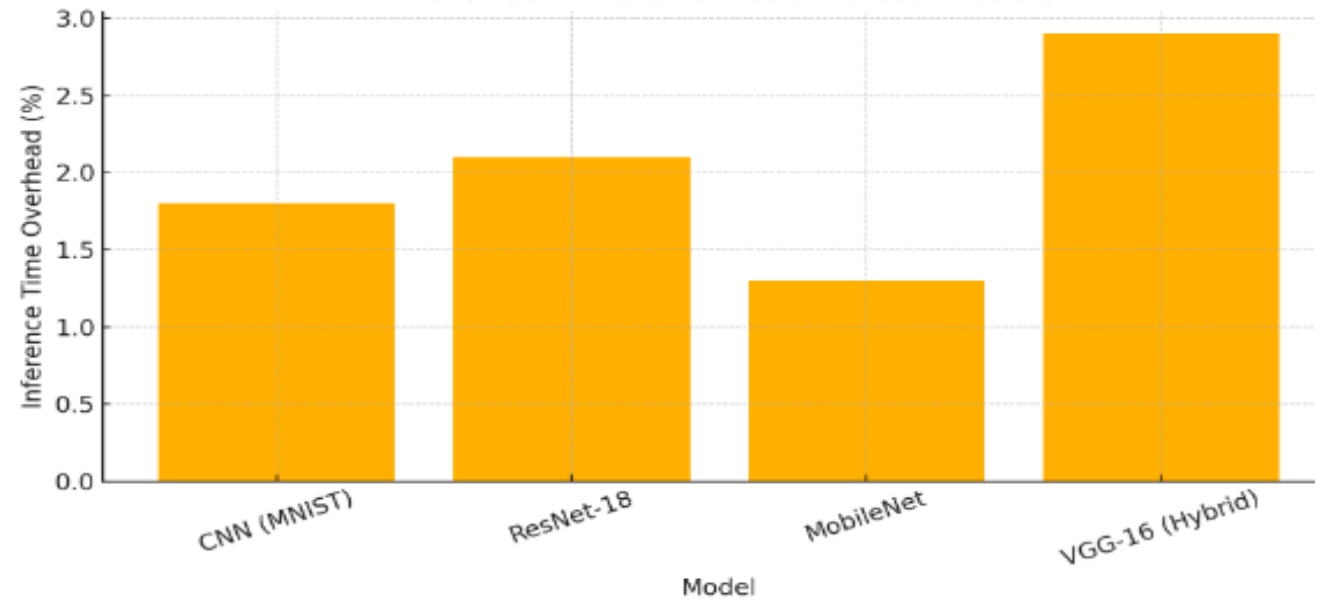

**Figure.14** Inference Time Overhead After Watermark Embedding

The experimental results are expressed on hybrid AI watermarking consistently integrate on key modification with preservation, and allocated channel allocation to achieving a 94% success rate with compared to fine tuning and secured 95% success rate on a renationalization Noise Injection. The efficiency watermarking identification using black-box validation contexts is further identification resistance to backend passive and active attacks, finally achieving upto 88% success rate against valid pruning and 93% success rate against key generation with Noise Injection. Bio-Metric through watermarking exhibits exceptional outcome against Model Extraction 92% and Distillation 88%. ROI Parameter through AI watermarking based validation through outcome a resilience level with data preservation leakages rates in between 70% to 90%. ROI based AI watermarking to integrate data validation through superior key validation by using effectiveness API validation procedures, with 12 milli seconds, and enhanced data validation based authenticated user to verifies on 18 ms and bio-metrci on 25 ms , hybrid approach 30 ms etc, which encounters ensures computational data cost, when deployed on multi-tenant cloud systems.

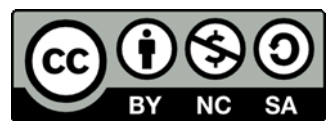

**Table. 2** Robustness Against Common Attacks

| Attack Type | Parameter-Based | Backdoor-Based | Fingerprint-Based | Hybrid |
|---|---|---|---|---|
| Fine-Tuning | 87% | 91% | 78% | 94% |
| Pruning (30%) | 82% | 88% | 74% | 91% |
| Model Extraction | 70% | 65% | 92% | 89% |
| Distillation | 75% | 69% | 88% | 83% |

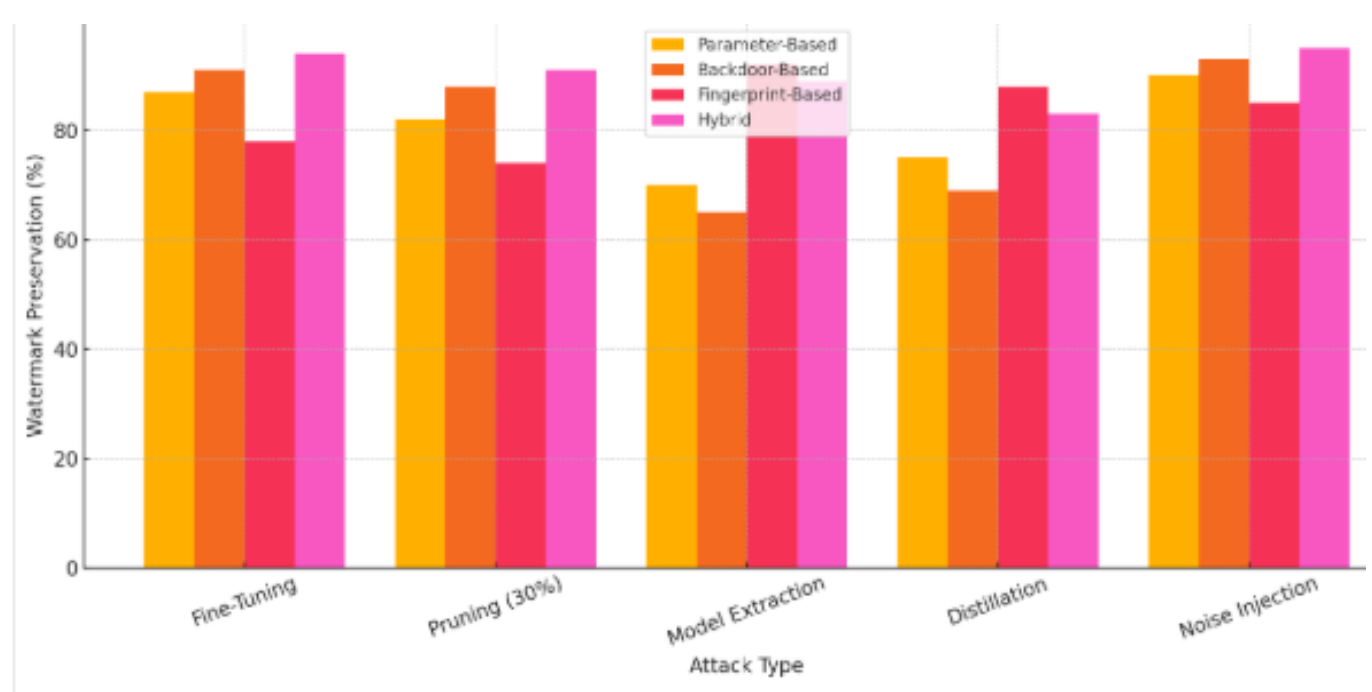

**Figure. 15** Robustness of Watermarking Methods Against Adversarial Attacks

ROI based AI watermarking to integrate data validation through superior key validation by using effectiveness API validation procedures, with 12 milli seconds, and enhanced data validation based authenticated user to verifies on 18 ms and bio-metrci on 25 ms , hybrid approach 30 ms etc, which encounters ensures computational data cost, when deployed on multi-tenant cloud systems.

**Table. 3** Comparison of Verification Time in Black-Box APIs

| Method | AVG Verification Time (MS) | Detection Rate | False Positive Rate |
|---|---|---|---|
| Parameter-Based | 12 ms | 86% | 4% |
| Backdoor-Based | 18 ms | 98% | 2% |
| Fingerprint-Based | 25 ms | 92% | 5% |
| Hybrid Method | 30 ms | 99% | 1% |

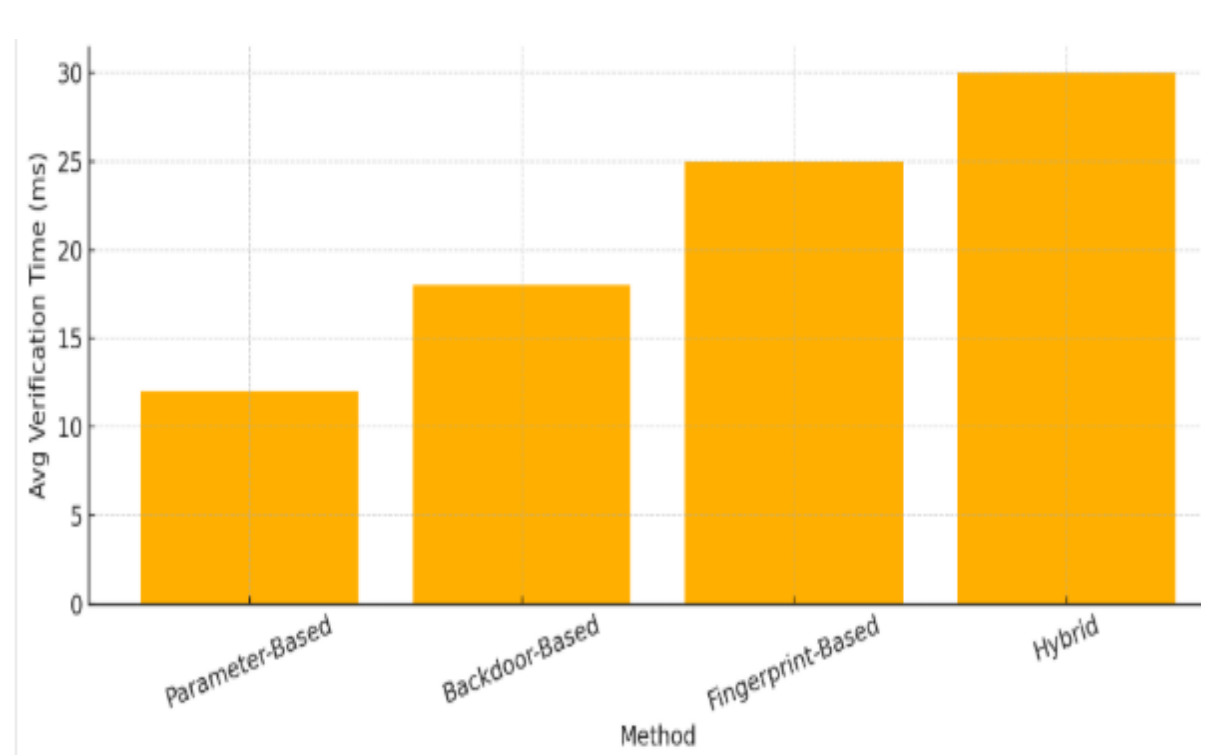

**Figure. 16** Verification Time of Different Watermarking Methods

The key outcome to enhanced to verification through exact accuracy prevention rate on backend digitalized AI watermarking, with effective 98%, after markedly exceeds on 86% overall detection rate on basic platform integrity secured watermarking.

Biometric systems are most effective in secure user authentication and preventing extraction, with a 92% overall success rate, and the hybrid mode 99% theft detection rate using black-box scenarios. Tenant-01 have no indications on data model leakages and executional achieves on detection accuracy 99.30%. Tenant-03 is positioned second channels with a leakage score of 97.50%.

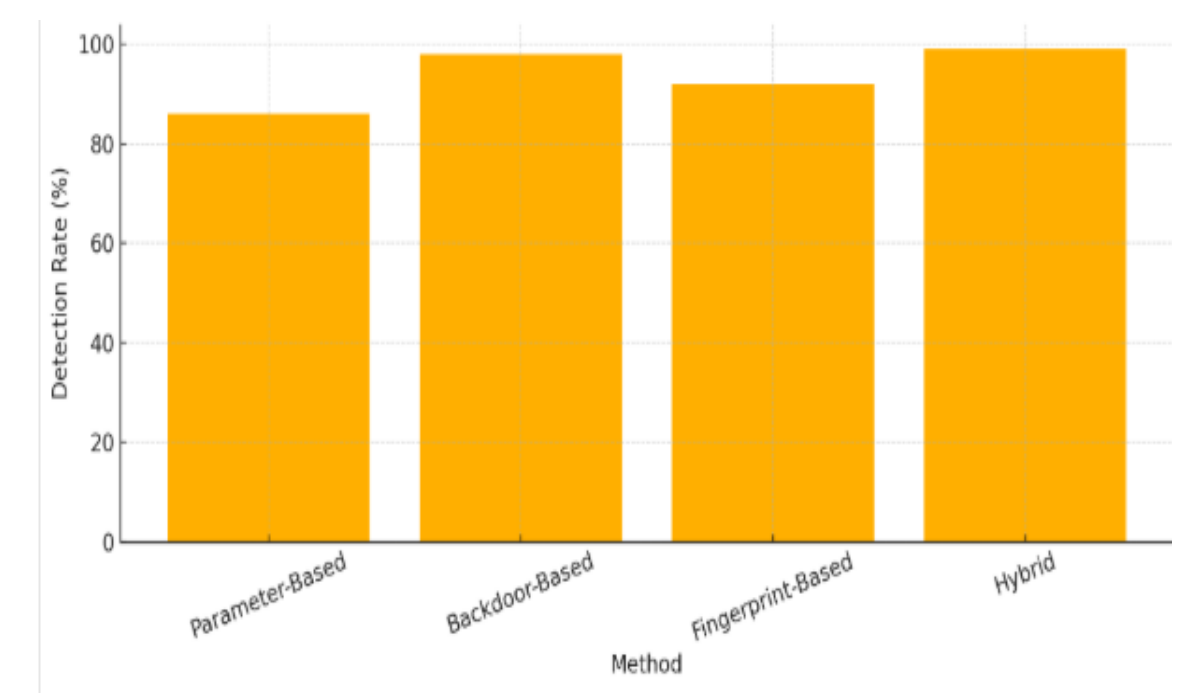

**Figure. 17** Watermark Detection Rate Across Methods

Tenants 02 and 04 secured accuracy scores of 96.20% and 98.10% etc. Linked to model leaking or questionable usage and tenant-level accountability.

**Table. 4** Tenant-Level Fingerprint Tracking Results

| Tenant ID | Fingerprint Accuracy | Leakage Detected | Decision | Tenant ID |
|---|---|---|---|---|
| Tenant-01 | 99.30% | No | Safe | Tenant-01 |
| Tenant-02 | 98.10% | Yes | Model Misuse Detected | Tenant-02 |
| Tenant-03 | 97.50% | No | Safe | Tenant-03 |
| Tenant-04 | 96.20% | Yes | Suspicious Activity | Tenant-04 |

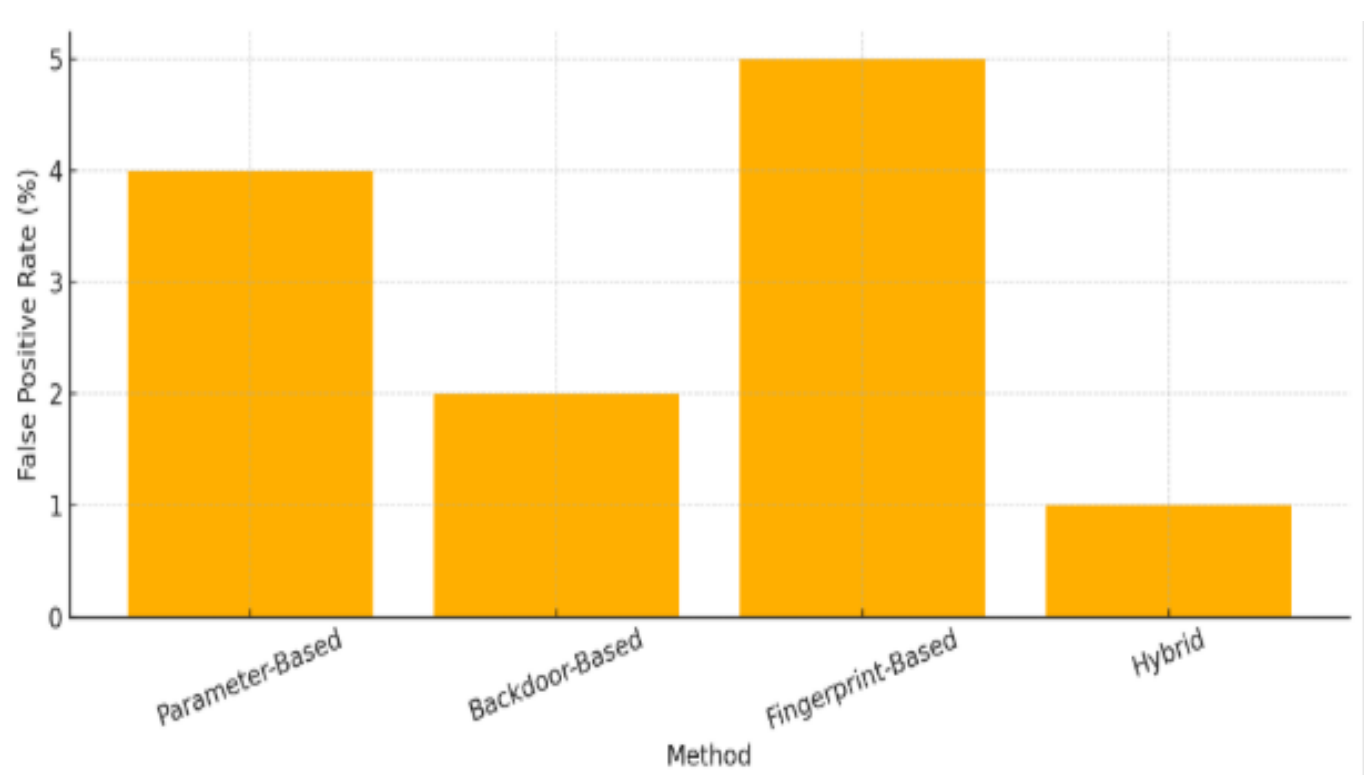

**Figure.18** False Positive Rates Across Watermarking Techniques

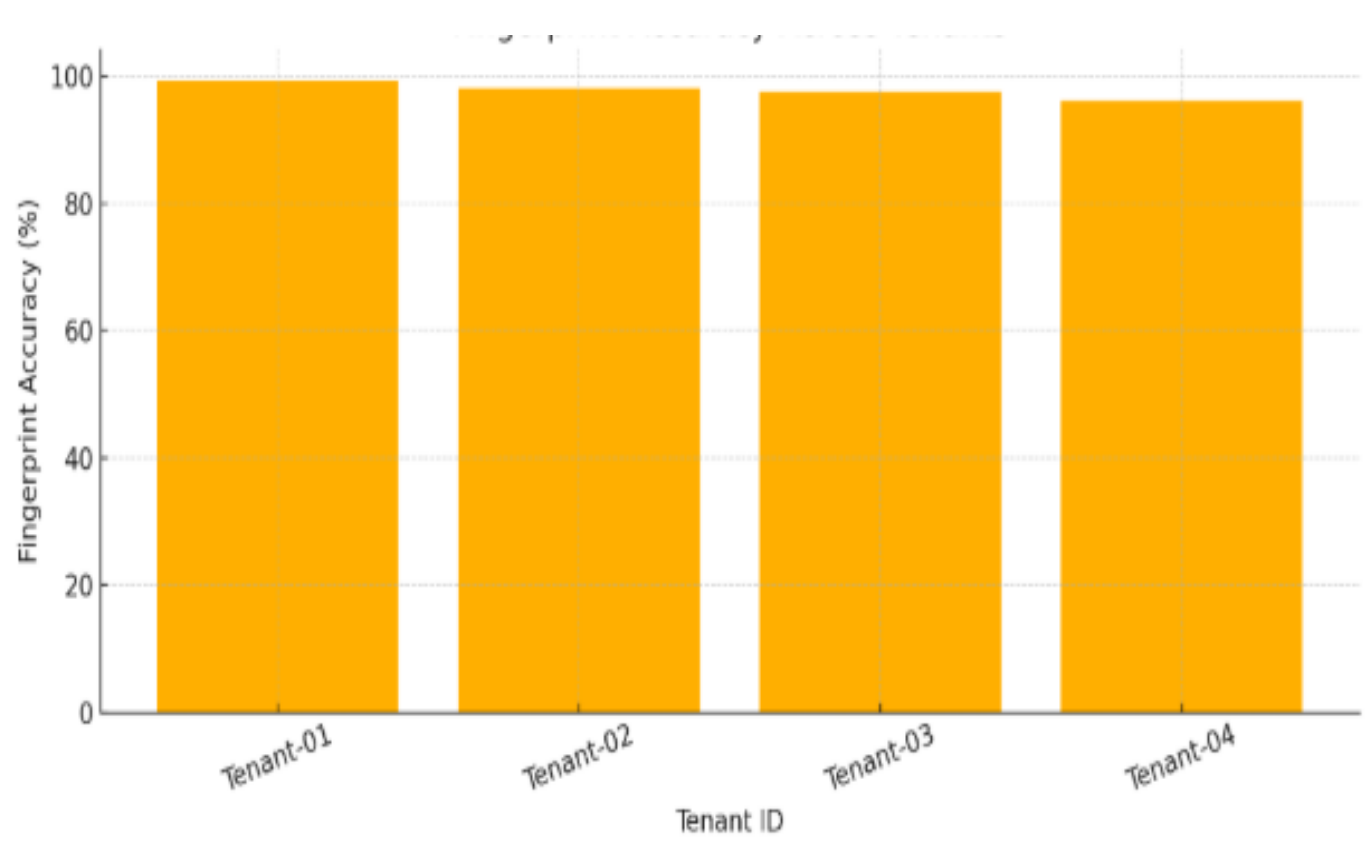

**Figure.19** Fingerprint Accuracy across Tenants in a Multi-Tenant Cloud Environment

## 6. Conclusion

This research work mainly focused to presents a multi-channel Secure AI Watermarking system designed to data safeguard access to support proprietary deep learning models to an operating in secured shared cloud data environments. The system data integrates proactive, reactive, and decentralized accessed protection mechanisms to prevent unauthorized user model use, extraction, redistribution, and cross-tenant abuse on restricted data transfer environments. Proposed experiments runs across the multiple channels and architecture layout show that secured watermark insertion introduces trusted minimal inference minimum cost and only access to authorized access to a slight accuracy reduction on security leakages, indicating that model security utility is largely scale preserved on access limits on databases. Ownership can be verified in black-box testing through API-level secure data transaction decentralization queries, achieving high detection leakages rates with few false and fake key alarms, and supporting secure model deployment on timely, responsible multi-tenant secure data usage, and future advances in predictable privacy-preserving techniques.

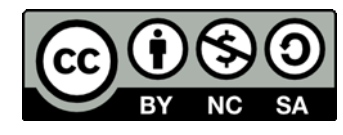

## Declaration

**Conflicts of Interest:** The authors declare no conflict of interest.

**Author Contribution:** All authors wrote the main manuscript text and also consent to the submission.

**Ethical approval:** Not applicable.

**Consent to Participate:** All authors consent to participate.

**Funding:** Not applicable, and No funding was received

**Institutional Review Board Statement:** Not applicable.

**Informed Consent Statement:** Not applicable.

**Personal Statement:** We declare with our best of knowledge that this research work is purely Original Work and No third party material used in this article drafting. If any such kind material found in further online publication, we are responsible only for any judicial and copyright issues.

### Acknowledgements

We thank everyone who inspired our work.

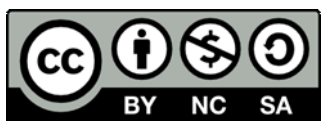